\documentstyle[BoxedEPSF]{article}
\begin{document}
{\bf Charge based quantum computer without charge transfer}

\bigskip
\bigskip
\noindent
V.V'yurkov and  L.Y.Gorelik$^{(1)}$\\

Institute of Physics and Technology RAS\\
Nakhimovsky prosp. 34, Moscow, 117218, Russia\\
Phone: (095)3324918, Fax: (095) 1293141\\
E-mail: vyurkov@ftian.oivta.ru\\

$^{(1)}$Dept. Applied Physics, Chalmers University of Technology\\
and Goteborg University, S-142 96, Goteborg, Sweden\\
Phone: +46 31-7723143, Fax: +46 31-416984\\
E-mail: gorelik@fy.chalmers.se

\bigskip
\it{
A novel implementation of a charge based quantum computer is proposed. There is no
charge transfer during calculation, therefore, uncontrollable entanglement between qubits
due to long-range Coulomb forces is suppressed. High-speed computation with 1ps per an
operation looks as feasible.}
\bigskip

\rm
Almost all recent experimental realizations of quantum computation accomplished on several qubits and even theoretical proposals of new
devices were based on spin states. Much cited papers of Kane \cite{1} and
DiVincenzo et al. \cite{2,3,4} just concerned nucleus or electron spin encoding
in solid state implementation of a quantum computer (QC). The
attractiveness of spin states in solid state~QC was mainly caused by quite
long decoherence time: hours for a nucleus spin and milliseconds for
electron spin seem attainable \cite{1,4}.

Nevertheless, space or charge states for quantum encoding do not seem less
prospective [5--13] although two main disadvantages are commonly
mentioned \cite{3}. Firstly, much less decoherence time is expected for charge
states. Secondly, charge transfer results in uncontrollable interaction
between even distant qubits due to long-range Coulomb forces. However, as
it was shown recently \cite{11,12}, for fairly small energy gap between
different charge (or space) states decoherence caused by acoustic phonons
might be very weak. 
Here we withdraw the second reproach for charge
based~QC offering a construction where no charge transfer occurs during
computation. Besides, the computation may be much faster than that in spin
based QC.
\newpage
{\bf Qubit and its operation}
\bigskip

A qubit is implemented in two double quantum dots (DQD) (Fig.1). A DQD
consists of a pair of quantum dots with a single electron. The electrode
$E$ operates on the strength of exchange interaction between DQDs. The
electrode $T$ varies tunneling coupling between quantum dots constituting a
DQD.

\HideDisplacementBoxes
\BoxedEPSF{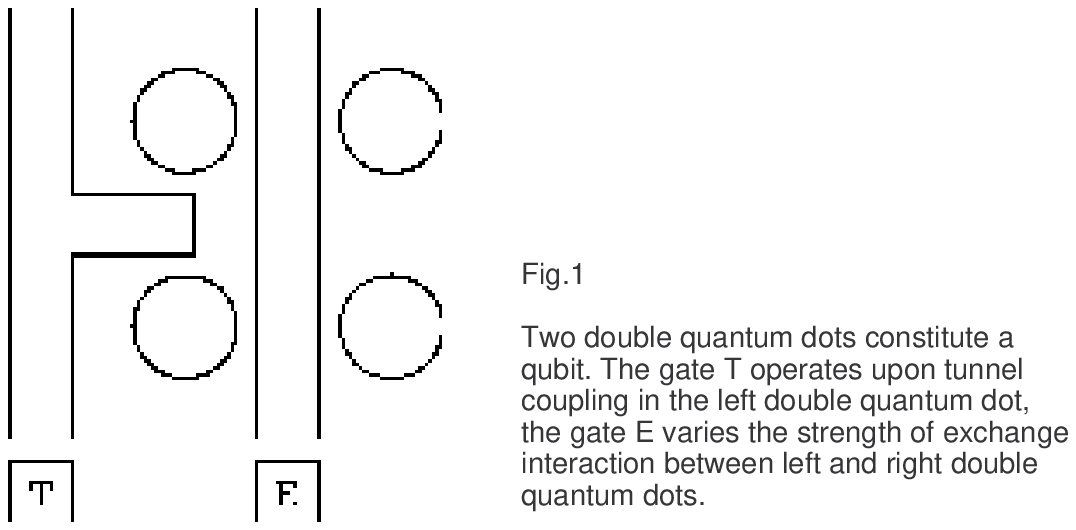}

We designate two lowest states of an electron in a DQD as $|+\rangle$
and $|-\rangle$. The state
$|+\rangle=\sqrt{1\over{2}}(\psi(r-r_1)+\psi(r-r_2))$
is symmetric and the state
$|-\rangle=\sqrt{1\over{2}}(\psi(r-r_1)-\psi(r-r_2))$ is anti-symmetric. Here $r_1$ 
and $r_2$  are coordinates of the
first and second quantum dots, $\psi(r)$ is an electron wave function in a dot, it decays as 
$e^{-{2r\over{a}}}$ 
outside the dot, the magnitude of $a$  depends on a potential barrier height
which may be controlled by a gate electrode voltage.

Two basic computational states of the qubit composed of two DQDs are
$$
|0\rangle=|+_-\rangle;\; \;\;|1\rangle=|_-+\rangle.
  \eqno{(1)}
$$
In these states an electron in the first DQD is in symmetric state and another electron in the
 second DQD is in anti-symmetric state, and vice versa.
\newpage
As an exchange interaction is to be involved later one should take into account the
Fermi-Dirac statistics of electrons as fermion particles. When two electrons are in the
triplet spin state (total spin $S=1$), i.\,e., have a symmetric spin configuration, their space
wave function is anti-symmetric with respect to permutation. Therefore,
instead of states~(1) one should use the following states:
$$
|0 >  = \sqrt {1\over{2}} \bigl(|+_1 -_2 \rangle  - |+_2  -_1  \rangle \bigr),
\eqno{(2)}
$$
$$
|1 >  = \sqrt {1\over{2}} \bigl(| -_1  + _2  \rangle  - | - _2  + _1 \rangle \bigr),
\eqno{(3)}
$$
where the indexes 1 and 2 correspond respectively to the first and second
electron.
Hereafter we assume an overall spin polarization of electrons in the system.

Any state of a qubit
$$
(|a|^2  + |b|^2 )^{ - {1\over{2}}} (a|0\rangle   + b|1\rangle)
\eqno{(4)}
$$
is created with the help of  gate electrodes. Here $a$ and $b$ are arbitrary complex numbers.
Thus qubit states belong to the Hilbert sub-space produced by orthogonal
states $|0\rangle$ and $|1\rangle$. Another pair of states $|++\rangle$
and $|--\rangle$ do not arise during qubit transformations. These
states could appear only due to decoherence processes. Worth noting that no charge
transfer is required to produce any of the states~(4). It can be easily confirmed that for any
numbers $a$ and $b$ the probability to find an electron in any quantum dot is constantly equal
to~0.5. This is just a reason why a qubit based on a pair of DQDs is required instead of a
qubit based on a single DQD.

Unitary transformations of a qubit are described by the Hamiltonian in matrix
presentation in basis of states  (2) and (3) 
$$
H = A\sigma_x  + P\sigma_z  = A
\pmatrix{
0 & 1  \cr
1 & 0}
+ P
\pmatrix{
   1 & \phantom{-}0  \cr
   0 &  - 1}.
\eqno{(5)}
$$
Here the factors $A$ and $P$ depend upon a voltage applied to gates $E$
and $T$. The associated
unitary time evolution is determined by the operator
$$
U(t) = \widehat T e^{i \int_{0}^{t}{ H(t')dt'\over{\hbar}}},
\eqno{(6)}
$$
where $\widehat T$ is a time ordering operator.
\newpage
In absent voltage regime the magnitudes $A$ and $P$ are supposed to equal zero. In reality,
their values less than $\hbar\over{\tau}$ where $\tau$ being a time of calculation are desirable. Otherwise, an
evolution of a qubit in an idle state should be put into account.

The amplitude shift can be realized by applying voltage to electrode $E$
which operates
upon the strength of exchange interaction between electrons located in surrounding DQDs
(Fig.~1). The applied positive voltage diminishes the potential barrier height and augments
the wave function overlap. It leads to enhancement of exchange interaction. In particular,
an amplitude flip resulting in a transition of a state $|0\rangle$ into
a state $|1\rangle$ can be performed by
means of appropriate pulse amplitude and duration. An exchange interaction is based on
Coulomb interaction between electrons
$$
U_C (r_1 ,\,r_2 ) = {e^2\over{\kappa |r_1  - r_2 |}},
\eqno{(7)}
$$
where $r_1$ and $r_2$  are respectively the coordinates of first and second
electron, $\kappa$  is a
permittivity of an environment. The direct calculation of matrix elements of
operator~$U_C$
for transitions from basic qubit states $|+_-\rangle$ and $|{}_-+\rangle$ to
complimentary states $|++\rangle$ or $|--\rangle$
gives zero in contrast with transitions between basic states. Obviously, beforehand the
states  $|++\rangle$ and $|--\rangle$  should be also presented in the form
like~(2)--(3). Moreover, the
transitions to these states are prohibited by symmetry conservation. Indeed,
$|++\rangle$ and
$|--\rangle$  states have additional symmetry with respect to inversion over
$E$ - electrode axis.
The operator $U_C$ can not break this symmetry.

A phase shift of a qubit is operated by the gate electrode $T$ acting on the
first DQD.
When a positive potential is applied tunneling between constituent quantum dots is
reinforced and the energy difference  $\Delta\epsilon$ between $|+\rangle$ and
$|-\rangle$ states of this DQD becomes
greater. It results in steadily rising phase difference between ${|+\rangle}$
and~${|-\rangle}$ states. In
particular, for a proper impulse duration and amplitude the phase difference
achieves $\pi$, i.\,e.
a phase flip occurs. It should be outlined that no tunneling can happen during a phase shift
operation as it needs a definite bias between quantum dots (see for the
section "Read-out").
\newpage
{\bf Two-qubit operations }
\bigskip

   $E$-electrodes placed between qubits make it possible to perform two-qubit operations.
The most simple to be realized is a SWAP operation, that is exchange of states between
neighbor qubits. This operation is fulfilled merely by sequential application of voltage to
electrodes $E$ resulting in exchange of states between surrounding DQDs. The SWAP
operation allows to move any qubit along a chain and put in contact any pair of qubits.

 Furthermore, the quantum XOR can be obtained by applying the sequence
of~$\pi\over{2}$ phase
shifts of a single qubit and a square root SWAP operation~[3]. Other 2-qubit
logic
operations (CNOT etc.) could be also composed in a similar way.

The rate of 1-qubit and 2-qubit operations depend on the strength of exchange interaction
and tunneling coupling augmented by voltage applied to electrodes. The offered
construction permits, in principle, a clock speed up to 1\,THz compared to
1\,GHz for elecron
spin based QC~[4] or 75\,kHz for nucleus spin based QC~[1].

The operational voltages are not less than of the order of potential barrier
height between
quantum dots. It is desirable to make them both lower.

\bigskip
{\bf Initialization of a computer}
\bigskip

In our opinion, the easy way of initial state creation in the QC under consideration is to
input electrons in a necessary state through the end of preliminary empty qubit chain. An
electron can be supplied to the extreme DQD by some single-electron device, for instance,
a turnstile. A symmetric state $|+\rangle$ can be prepared by applying a fairly high voltage to an
electrode $T$ to make a sufficient energy difference between $|+\rangle$ and $|-\rangle$
states. The first one is a ground
state of an electron in a DQD. Thus for fairly large energy
difference between
ground $|+\rangle$ and excited $|-\rangle$ states with respect to a thermal energy kT the occupancy of the
upper state can be done negligible. A~state $|+\rangle$ can be easily
inverted into a state $|-\rangle$ (if
required) when a DQD is biased by means of some additional electrode placed near one
quantum dot. Afterwards, an electron with a formed state can be pushed from the starting
DQD to any DQD in a chain by voltage pulses successively applied to $E$-electrodes along
the path. In this way one can pump electrons one by one along a qubit chain and fill all
DQDs.

All electrons in the proposed QC should be spin-polarized as an exchange interaction
depends on a spin configuration. Two possibilities to make it seem like plausible.
Obviously, it could be done by external magnetic field and cooling. Another way is to
supply electrons from a ferromagnetic particle.
\newpage
There is a possibility to speed up $|+\rangle$  or $|-\rangle$  states
formation even under condition $\Delta\epsilon<kT$
when a thermal relaxation is evidently ineffective. The procedure is merely
the reversed read-out considered in the next section.

\bigskip
{\bf Read-out}
\bigskip

To read out the information accumulated in the resultant register of a QC one should
measure the state of DQDs. The following procedure is proposed. The voltage applied to
the gate $T$ makes an essential energy gap $\Delta\epsilon$  between $|+\rangle$
and $|-\rangle$ states. The mean energy
of a system when electron occupies one quantum dot is situated just in the middle of the
gap. When this DQD is biased by a voltage $\Delta\epsilon\over{2e}$ by means of outer electrode placed near
one quantum dot resonant tunneling of an electron to a quantum dot occurs. To what dot of
two depends on initial state. Surely, the state when an electron is located in one quantum
dot is not an eigen state of a system. Actually on applying a bias the electron begins to
oscillate between quantum dots. But what dot it visits first depends on the initial state of a
DQD, whether it is $|+\rangle$ or $|-\rangle$. One could switch off $T$-voltage,
i.\,e. tunneling, right at the

moment when electron is located in one quantum dot. Measuring a conductance of a
quantum wire placed nearby one can easily distinguish what quantum dot contains the
electron~[5]. Moreover, a current-voltage curve of a single wire placed along a resultant
qubit chain could, in principle, provide the whole information about its
charge state~[8].

\bigskip
{\bf Decoherence: voltage fluctuations and phonons}
\bigskip

The weaker is decoherence the longer is fault tolerant time of a QC. The decoherance in
our QC circuit arises when a state of some DQD is altered from $|+\rangle$ to
$|-\rangle$, or vice versa.
Two comprehensible sources of decoherence in the system are voltage fluctuations and
phonons.

Voltage fluctuations were already discussed in~[1]. Resonant transitions between
states there were induced by associated spectrum component of a voltage noise. In the
present construction voltage fluctuations on electrodes $T$ and $E$ can not affect transitions
between $|+\rangle$ and $|-\rangle$ states owing to a symmetry of the structure. However, noise to signal
ratio influences on the accuracy of 1-qubit and 2-qubit operations.

In spin based QCs the voltages required for manipulation with a qubit shift the frequency
of NSR or ESR of nearby qubits. This is why the immediate operation on distant qubits is
only possible unless necessary corrections are not undertaken. One more advantage
relevant to offered implementation is that manipulation with some qubit does not perturbs
the state of neighbor ones.
\newpage
Anyway, optical gates to control a potential barrier transparency~[5--7,\,13]
could be used
instead of voltage gates. Extremely strong dependence of a photo-stimulated tunneling on a
beam polarization permits a precise addressing to a definite qubit.

Were voltage fluctuations suppressed the action of phonons did not cease. Nevertheless,
cheerful results were obtained recently in~[11,\,12]. They much contradict with
intuitive expectations. Here we restrict the discussion by rather qualitative consideration.
The states $|+\rangle$ and $|-\rangle$ have very small energy difference
$\Delta\epsilon=A$ in an idle qubit. The same
is the energy of an acoustic phonon $\epsilon$ required to enforce a transition between these states.
The lower is the energy the smaller is the matrix element of these transitions. One reason is
a long wave length of a relevant phonon. In~[12] a probability $1\over{\tau}$
of spontaneous
emission of acoustic phonons in DQD was calculated. There were obtained dependencies 
$\tau\sim\epsilon^{-5}$ for deformation acoustic phonons and $\tau\sim\epsilon^{-3}$
for piezoelectric acoustic phonons. The
latter do not exist in Si but dominate in $A_3B_5$ compounds. When two GaAs quantum dots
constituting a DQD are separated by a distance $r=22\,nm$ the values
$\tau\sim 10^{-2}$\,s for
piezoelectric acoustic phonons and $\tau\sim 10^{6}$\,s for deformation
acoustic phonons are attainable.
The stimulated emission and absorption, and multi-phonon processes were beyond the
frame of~[11,\,12]. However, they may be significant when $\epsilon<kT$.
To evaluate the
stimulated emission one should multiply the above magnitude of~$\tau$~by 
a small factor ${\epsilon}\over{kT}$
originating from the Bose-Einstein statistics for $\epsilon\ll kT$
$n(\epsilon) = {1\over{e^{\epsilon/kT} - 1}} \approx\
{kT\over{\epsilon}}$.
Nevertheless, the probability to emit or to absorb a phonon still drops with
energy $\epsilon$. It
looks amazing as one can sustain the coherence between states separated by energy less
than $kT$ for an arbitrary long time.

In our opinion, in reality two-phonon processes set a limit for decoherence time.
During
transition from a state $|+\rangle$ to a state $|-\rangle$ one phonon is
emitted and another absorbed. The
energy of phonons involved is about $kT$, that is, independent on energy
split $\Delta\epsilon$ between
states $|+\rangle$ and $|-\rangle$.

\bigskip
{\bf Fabrication}
\bigskip

As almost all so far offered solid state implementations of a QC the present one is also
based on tunneling operated by gate voltage. To reach the clock frequency 1\,THz not less
must be a tunneling rate. Thus the distances between quantum dots or dopant atoms should
be about tens nanometers. This size is beyond the state of art in modern technology.
Evidently, one could sacrifice speed in favour of greater dimensions. Photo-stimulated
tunneling instead of voltage-stimulated one may simplify demands for technology.
\newpage
In conclusion, a novel implementation of a charge based quantum computer is proposed.
A qubit consists of four quantum dots with two electrons. Evolution of the system is
controlled with gate voltages operating on tunneling coupling and strength of exchange
interaction. There is no charge transfer during calculation, therefore, uncontrollable
entanglement between qubits due to long-range Coulomb forces is suppressed. High-speed
computation and long fault tolerant time look as feasible.

\bigskip
{\bf Acknowledgments}
\bigskip

   The work was supported by the programs ``Prospective technologies and devices of
micro- and nanoelectronics'' and ``Physics of solid state nanostructures''
of the Russian
Ministry of Science and by the Swedish KVA. One of the authors (V.\,V.)
is much thankful
to L.\,Fedichkin for fruitful discussions.

\end{document}